\documentstyle[12pt]{article}
\begin{document}

%
%

\title{The Second Moment of the Structure Function
          for Pseudoscalar Mesons in QCD Sum Rules\\}

\author{ Yoshinori Nishino\\ \\
  \small{\it Research Center for Nuclear Physics,} \\
  \small{\it Osaka University, Ibaraki, Osaka 567, Japan}\\ }

\date{  }

\maketitle

%
%

\begin{abstract}
We calculate the values of the second moment $M_2^s$ of
the flavor-singlet structure function $F_2^s$ for the pion and
the kaon in QCD sum rules, and investigate how the results depend
on their flavor structure (quark contents). Our calculations give similar
values of $M_2^s$ for these two mesons, because of cancellation among
several non-small factors. We emphasize that decay constants,
meson masses and quark masses play an essential role in the
above cancellation.
\end{abstract}

\newpage

%
%

Structure functions carry information on dynamical distributions
of quarks and gluons inside hadrons. Intensive studies of nucleon structure
functions have been performed  using various lepton induced reactions.
In contrast, those of other hadrons are not easy to access.
It is not yet clear how the difference in quark contents of hadrons
affects the structure functions. In the present paper, we would like
to study this problem by calculating the values of the second moment
$M_2^s$ of the flavor-singlet structure function $F_2^s$ for the pion
and the kaon in QCD sum rules (QSR)

The QSR method, developed by Shifman, Vainshtein and Zakharov
\cite{svz}, has been used to study a number of problems in hadron physics
\cite{rep}. Belyaev and Ioffe studied the Bjorken variable dependence of the
structure function $F_2(\xi,Q^2)$ of the nucleon based on QSR,
and showed that their method is applicable only in the intermediate region
of the Bjorken variable $\xi$ \cite{bioffe}. On the other hand, moments
of structure functions can be studied by introducing constant external
fields into QSR \cite{Ioffes}. Using this technique, Belyaev and Blok
calculated $M_2^s$ values for the pion and the nucleon \cite{bb}.
There also exist some applications of this technique to moments of
other structure functions \cite{sf}. These calculations indicate that
QSR is a useful tool in calculating moments of structure
functions. Below we adopt this technique to investigate $M_2^s$ values
for the pion and the kaon, and study in particular the effects of decay
constants, meson masses and quark masses on the momentum
fraction carried by quarks.

%
%

The second moment $ M^s_2(\mu^2)$ of the (unpolarized) flavor-singlet
structure function $F_2^s$ is defined as follows:
\begin{equation}
       M^s_2(\mu^2)p_\mu p_\nu
        =\frac{1}{2}\langle  \mbox{hadron}(p) |  \Theta^q_{\mu \nu}
       | \mbox{hadron}(p) \rangle .\label{eq:def-m2}
\end{equation}
Here $| \mbox{hadron}(p) \rangle $ is the momentum eigenstate of
the hadron concerned, $\mu$ is the renormalization point of
the operator, and $\Theta^q_{\mu \nu} $ denotes the quark part of
the symmetrized energy-momentum tensor,
\begin{equation}
   \Theta^{q}_{\mu \nu}(x) \equiv \frac{\mbox{i}}{4} \left(
             \bar{\psi}(x)\stackrel{\leftrightarrow}{D}_{ \mu }
             \gamma _{ \nu } \psi (x)
            +\bar{\psi}(x)\stackrel{\leftrightarrow}{D}_{ \nu }
             \gamma _{ \mu} \psi (x) \right) ,\label{eq:theta-q}
\end{equation}
where $\psi(x)$ and $D_\mu$ are the quark field and the
covariant derivative,  respectively.
$M_2^s$ can also be represented in terms of (unpolarized) quark
distribution functions $f_q(x,\mu^2)$:
\begin{equation}
            M_2^s(\mu^2) = \int_0^1 \mbox{d}\xi \; \xi
            \; \sum_q [f_{q}(\xi,\mu^2)+
            f_{\bar{q}}(\xi,\mu^2)]. \label{eq:def'-m2}
\end{equation}
$M_2^s$  can thus be interpreted as the momentum fraction
carried by quarks.

In QSR, hadronic matrix elements of local operators, like $M_2^s$,
can generally be calculated by introducing constant external fields
\cite{Ioffes}.
In the present case, the external field has the tensor structure
corresponding to the operator $\Theta^{q}_{\mu \nu}$.
\begin{equation}
       \Delta {\cal L} (x) =  -\Theta^q_{\rho \lambda}(x) S^{\rho\lambda}.
\end{equation}
Under existence of this external tensor field, we consider the
two-point function of the interpolating field for the hadron concerned.
For pseudoscalar mesons, appropriate interpolating fields
are axial vector currents.
\begin{equation}
    j_{\mu}^5(x)=
 	\left\{
    	\begin{array}{lll}
 	   &\bar{u}(x)\gamma_\mu\gamma_5 d(x)
                                  \hspace{1cm} &\mbox{for the pion},\\
    	    &\bar{u}(x)\gamma_\mu\gamma_5 s(x)
                                                        &\mbox{for the kaon},
	\end{array}
	\right.
\end{equation}
Thus the two-point function in the present case is given by
\begin{equation}
       \Pi_{\mu\nu}(Q^2) = \mbox{i} \int \mbox{d}^4x \, \mbox{e}^{\mbox{i}qx}
	   \langle  T[j_{\mu}^5(x), j_{\nu}^{ 5 \dagger}(0)]
	   \rangle _S \hspace{0.5 cm}
	    (Q^2=-q^2),\label{eq:pi-m2}
\end{equation}
where the subscript "$S$" represents existence of the external field.
{}From Eq.(\ref{eq:pi-m2}), we extract the first order term
$\Pi_{\mu\nu\rho\lambda}^1(Q^2) $
in the external field $S^{\rho\lambda}$ ($\Pi_{\mu\nu}(Q^2)=
\Pi_{\mu\nu}^0(Q^2) + \Pi_{\mu\nu\rho\lambda}^1(Q^2)
S^{\rho\lambda}$+\ldots).

According to  the QSR method, we calculate
$\Pi_{\mu\nu\rho\lambda}^1(Q^2) $ in two different ways.
In the phenomenological side of QSR, we represent
$\Pi_{\mu\nu\rho\lambda}^1(Q^2) $ in terms of $M_2^s$, using
the reduction formula and the dispersion relation.
On the other hand, in the operator product expansion (OPE)  side of QSR,
we expand $\Pi_{\mu\nu\rho\lambda}^1(Q^2) $ into terms with
various condensates. In the region of $Q^2$ where these two
calculations are manageable, we can equate both sides of QSR and
can obtain QSR for $M_2^s$. In actual calculations we perform
the Borel transformation on both sides of QSR.

First, we shall consider the phenomenological side of QSR.
Hereafter we show calculations only for the kaon, since
expressions for the pion are obtained in the same way.
With the help of the reduction formula and the double dispersion relation,
we get
\begin{eqnarray}
\Pi_{\mu\nu\rho\lambda}^1(Q^2)
        &=&-4f_K^2\frac{M_2^s}{(Q^2+m_K^2)^2}
             q_\mu q_\nu q_\rho q_\lambda
            +\sum_{i}\frac{(B_{i})_{\mu\nu\rho\lambda}}
              {(Q^2+m_K^2) (Q^2+m_{i}^2)}\nonumber \\
              &&\mbox{}+
             \sum_{ij}\frac{(C_{ij})_{\mu\nu\rho\lambda}}
              {(Q^2+m_{i}^2)(Q^2+m_{j}^2)} \nonumber \\
              &&\mbox{}
              +\Pi_{\mu\nu\rho\lambda}^1
                (Q^2 > s_0: \mbox{perturbative}), \label{eq:rf-pi}
\end{eqnarray}
where $B_i$ and $C_{ij}$ represent matrix elements which
are not diagonal in the kaonic state, and $m_i$ is the mass of
the $i$-th excited state in the axial vector channel. And we
assumed that perturbative calculations are correct in $Q^2$
above $s_0$.

In Eq.(\ref{eq:rf-pi}), $M_2^s$ is appearing as the coefficient of
$q_\mu q_\nu q_\rho q_\lambda$. So we extract the coefficient of
$q_\mu q_\nu q_\rho q_\lambda$ from
$\Pi_{\mu\nu\rho\lambda}^1(Q^2)$
and then apply the Borel transformation $\hat{L}_M$.
\begin{equation}
                   \hat{L}_M \equiv \lim_{Q^2,n \to \infty \atop
                   \frac{Q^2}{n}=M^2}\frac{1}{(n-1)!}
                   (Q^2)^n(-\frac{\mbox{d}}{\mbox{d}Q^2})^n ,
\label{eq:borel-def}
\end{equation}
where $M$ is the Borel mass. In this way, we finally obtain
\begin{equation}
                \hat{L}_M\Pi'^1(Q^2) =
                (-4f_K^2\frac{M_2^s}{M^4}+\frac{1}{M^2}b) \;
                \mbox{e}^{-\frac{m_K^2}{M^2}}
                +\hat{L}_M\Pi'^1(Q^2> s_0: \mbox{perterbative}),
\label{eq:right-m2}
\end{equation}
as the phenomenological side of QSR.
Here $b$ is a constant which comes from the second term of
Eq.(\ref{eq:rf-pi}) and we neglected all exponentially small
terms compared to the first term in the r.h.s. of
Eq.(\ref{eq:right-m2}).

On the other hand,  in the OPE side of QSR we apply OPE to the
T-product in the integrand of Eq.(\ref{eq:pi-m2}).
To be consistent with the phenomenological side,
we consider only the coefficient of
$q_\mu q_\nu q_\rho q_\lambda$, and
then perform the Borel transformation. After some calculations
we obtain,
\begin{eqnarray}
          \hat{L}_M \Pi'^1(Q^2) &=& -\frac{1}{2\pi^2}\frac{1}{M^2}
                   -\frac{1}{18}\langle \alpha_s FF\rangle
                      \frac{1}{M^6}
                  -2 \frac{m_u\langle \bar{u}u\rangle
                       +m_s\langle \bar{s}s\rangle }{M^6}
                  \nonumber \\
               &&\mbox{}
                  +\frac{1}{3} \frac{m_u\langle \bar{u}
                      \sigma_{\mu\nu}F^{\mu\nu}u\rangle
                      +m_s\langle \bar{s}\sigma_{\mu\nu}
                      F^{\mu\nu}s\rangle}{M^8} \nonumber \\
              &&\mbox{}
                    -\frac{16\pi}{27}\alpha_s
                      \frac{\langle \bar{u}u\rangle ^2+\langle
                      \bar{s}s\rangle ^2}{M^8}
                    +\frac{64\pi}{27}\alpha_s\frac{\langle
                      \bar{u}u\rangle \langle \bar{s}s\rangle }{M^8}
                    \label{eq:left-m2}.
\end{eqnarray}
Here we calculated up to dimension-6 condensate terms and the first
order in quark masses, whereas in Ref.\cite{bb} they neglected quark
masses. Wilson coefficients are obtained by calculating diagrams
depicted in Fig.1. Note that the diagrams in which the external field
interacts with vacuum quarks contribute to dimension-6
condensate terms, but they are small compared to the other
dimension-6 condensate terms \cite{bb}. The smallness of this term
is intuitively understandable, since low momentum quarks have less
contribution to the momentum fraction.

\marginpar{ \hspace{1cm} Fig.1}

Equating Eq.(\ref{eq:right-m2}) and Eq.(\ref{eq:left-m2}),
we can get QSR for $M_2^s$. But beforehand we have to
 perform the  QCD evolution to get $M_2^s$ for arbitrary
 renormalization point. After the QCD evolution, we get QSR
 for $M_2^s(\mu^2)$
\begin{eqnarray}
          &&\mbox{} \frac{9}{25}(1-L^{50/81})
             +L^{50/81}\frac{\mbox{e}^{m_K^2/M^2}}{f_K^2}
                \left[\frac{1}{8\pi ^2}M^2(1-\mbox{e}^{-s_0/M^2})
                \right.\nonumber \\
          &&\mbox{}+\frac{1}{72}
                 \frac{\langle \frac{\alpha _s}{\pi }FF\rangle }{M^2}
               +\frac{1}{2}
                 \frac{m_u\langle \bar{u}u\rangle +
                 m_s\langle \bar{s}s\rangle }{M^2}
                \nonumber \\
          && \mbox{}  -\frac{1}{12}
                \frac{m_u\langle \bar{u}\sigma_{\mu\nu}
                F^{\mu\nu}u\rangle
              +m_s\langle \bar{s}\sigma_{\mu\nu}
                F^{\mu\nu}s\rangle }{M^4}\nonumber \\
          &&\mbox{} \left.
              +\frac{4\pi}{27}\alpha_s
                \frac{\langle \bar{u}u\rangle ^2+
                \langle \bar{s}s\rangle ^2}{M^4}
             -\frac{16\pi}{27}\alpha_s
                \frac{\langle \bar{u}u\rangle \langle \bar{s}s\rangle}
                {M^4}\right] \nonumber \\
           &=& M_2^s(\mu^2)+C M^2
    \label{eq:borell-m2},
\end{eqnarray}
where
\begin{equation}
	L=\ln(\frac{M^2}{\Lambda^2})/\ln
	    (\frac{\mu^2}{\Lambda^2})
	\nonumber
\end{equation}
and $C$ is a constant.
In Eq.(\ref{eq:borell-m2}), we moved the contribution from
the continuum to the l.h.s. Note that four-quark condensate terms are
slightly different from those previously obtained by Belyaev and
Blok in the pion case \cite{bb}.

For a given $\mu^2$, the l.h.s. of Eq.(\ref{eq:borell-m2}) is
a linear function of the Borel mass squared $M^2$ with
$M_2^s(\mu^2)$ as a constant term, whereas the l.h.s. of
Eq.(\ref{eq:borell-m2}) is a very complicated function of $M^2$.
We approximate the latter by a linear function in the region of
$M^2$ where QSR is applicable. This region of $M^2$ is obtained
according to the  following condition; (i) the order of
the contribution from the highest order term of OPE is less than
10 \% of the l.h.s., and (ii) contribution from the continuum is
less than 50 \% of the other. In this way, we calculated
$M_2^s(\mu^2)$ for various values of the continuum threshold
$s_0$, and find stable $s_0$, which has the least influence
on the result.

%
%

We plotted the r.h.s. of Eq.(\ref{eq:borell-m2}) in Fig.2
for both the pion and the kaon cases. Here the values of the continuum
thresholds are $s_0=0.8\mbox{GeV}^2$ for the pion and
$s_0=1.2\mbox{GeV}^2$ for the kaon.

\marginpar{ \hspace{1cm} Fig.2}

We thus obtain the results,
\begin{eqnarray}
         M_2^s(\mu^2=49{\mbox{GeV}}^2)&=&0.39 \pm 0.04
         \hspace{1cm} \mbox{for the pion},\label{eq:result-pi}\\
         M_2^s(\mu^2=49{\mbox{GeV}}^2)&=&0.41 \pm 0.04
         \hspace{1cm} \mbox{for the kaon}, \label{eq:result-k}
\end{eqnarray}
Here, for numerical calculations, we used following values:
$ \langle\frac{\alpha_s}{\pi}FF\rangle =
	1.2\times 10^{-2}\mbox{GeV}^4 $,
	$ \langle\bar{s}s\rangle = 0.8\langle\bar{q}q\rangle $ \.
		($q$: $u$ or $d$) ,
	$ m_q\langle\bar{q}q\rangle = -(0.096\mbox{GeV})^4 $ ,
	$ m_s\langle\bar{s}s\rangle = -(0.21\mbox{GeV})^4 $ ,
 $ \langle \bar{s}\sigma_{\mu\nu}
            F^{\mu\nu}s\rangle = 0.8\langle \bar{q}\sigma_{\mu\nu}
            F^{\mu\nu}q\rangle $ ,
	$ \langle \bar{q}\sigma_{\mu\nu}
            F^{\mu\nu}q\rangle = 2m_0^2\langle\bar{q}q\rangle $ ,
	$ m_0^2 = 0.4 \mbox{Gev}^2 $ ,
 $ \alpha_s\langle\bar{q}q\rangle^2 = 1.8\times 10^{-4}
            	                      \mbox{GeV}^6	$ ,
 $ \Lambda = 150 \mbox{MeV}$,
 $ f_\pi=93 \mbox{MeV} $  and
 $ f_K=114 \mbox{MeV} $.

Our result in the pion case can be compared with the following values.\\
(i) NLO analysis of the Drell-Yan data \cite{m2-exp}
\begin{equation}
           2\int_0^1 \mbox{d}\xi \; \xi f_{\mbox{\scriptsize valence}}
           (\xi,\mu^2=49 \mbox{GeV}^2)=0.40\pm 0.02.
\end{equation}
(ii) Lattice calculation \cite{m2-lattice}
\begin{equation}
           M_2^s(\mu^2=49 \mbox{GeV}^2)=0.46 \pm 0.07.
\end{equation}
They are consistent with our result.

Next we discuss about the flavor (quark contents) dependence of
structure functions. Though the two values, (\ref{eq:result-pi})
and (\ref{eq:result-k}), are close to each other, it does not imply
that their dynamical origins are similar, as we can see below.
Since the sum rules for these two mesons are the same
in their form, differences are in the input values
of decay constants, meson masses and quark mass dependent terms
($m_q\langle\bar{q}q\rangle$ and $m_s\langle\bar{s}s\rangle$).

First to see how the difference in decay constants affect the
$M_2^s$ values, we put masses of mesons and quarks
to be zero (chiral limit). Then we have 0.38 as the $M_2^s$ value
for the pion, and 0.34 as that for the kaon.
The fact that in the chiral limit the $M_2^s$ value for the kaon is
smaller than that for the pion is easily understood from the sum rule
Eq.(\ref{eq:borell-m2}).
The difference in these two calculations comes from the values of
decay constants, where larger decay constant for the kaon
results in smaller $M_2^s$ value than that for the pion.

The roles of meson masses and of quark masses are understood
in the following manner.
In the pion case,  even if we put the pion mass and/or quark masses
to be zero,  the result is almost unchanged.  It can be seen by comparing
the chiral limit result 0.38 with 0.39 of Eq.(\ref{eq:result-pi}).
Therefore the meson/quark masses play a minor role in the calculations
for the pion.

However the situation is quite different in the kaon case.
The finiteness of the meson and quark masses
(the mass effect) lift the value of $M_2^s$ for kaon
from 0.34 (chiral limit) up to 0.41 of Eq.(\ref{eq:result-k}).
Moreover each of the meson mass and the quark mass has
much larger effect on the $M_2^s$ value for the kaon.

Since the meson mass is appearing in Eq.(\ref{eq:borell-m2}) as
$\exp(m_K^2/M^2)$, larger meson mass results in larger $M_2^s$.
And if we simply neglect the kaon mass, we have 0.26 as
$M_2^s$ for the kaon.
On the other hand, if we simply neglect terms with
quark masses, we have 0.51 as $M_2^s$ for the kaon.
Though these two effects are quite large, their directions are opposite
to each other.  Thus their partial cancellation gives the forementioned
mass effect.

%
%

In summary, we calculated  $M_2^s$ values
for the pion and the kaon in QCD sum rules,
and obtained similar values of $M_2^s$ for these mesons.
The value of $M_2^s$ for the pion is consistent with the lattice
calculation and with the experimental analysis.

We then discussed about the flavor (quark contents) dependence of
structure functions.
 Though larger decay constant for the kaon makes
$M_2^s$ decrease compared to that for the pion,
this effect almost cancels the mass effect.
Also, each of the kaon mass and quark masses
has a large effect on $M_2^s$ for the kaon, and
non-zero values of them are essential.

%
%
\vspace{.3cm}

The author would like to thank T. Suzuki for suggestion of the problem, and
him and S. Tadokoro for valuable discussions and encouragement.
Thanks are also due to E. G. Drukarev for a useful advice.

%
%

\newpage
\subsection*{Figure Caption}

{\bf Fig.1}\\
        Typical diagrams that contribute to $\Pi'^1(Q^2)$.
        The wavy lines and the curry lines represent the external
         tensor fields and the background gluon fields, respectively.
\vspace{0.5cm} \\
{\bf Fig.2}\\
      The solid lines indicate the OPE side of QSR at
      $\mu^2=49\mbox{GeV}^2$.
       We approximate the solid lines by linear lines in the region
      of $M^2$ in between the dotted lines.


\begin{thebibliography}{99}
    \bibitem{svz}
   M. A. Shifman, A. I. Vainshtein and V. I. Zakharov,
    Nucl.  Phys.  {\bf B 147} (1979) 385, 448, 519.
    \bibitem{rep}
   L. J. Reinders, H. Rubinstein and S. Yazaki,
    Phys. Rep. {\bf 127} (1985) 1.
   \bibitem{bioffe}
     V. M. Belyaev and B. L. Ioffe,
     Nucl. Phys. {\bf B 310} (1988) 548;
     {\bf B313}(1989)647;\\
     J.P. Singh and J. Pasupathy,
     Phys. Lett. {\bf 198 B} (1987) 239.
      \bibitem{Ioffes}
         B. L. Ioffe and A. V. Smilga,
         Nucl. Phys. {\bf B 232} (1984) 109.
      \bibitem{bb}
          V. M. Belyaev and B. Yu. Blok,
          Phys. Lett. {\bf 67 B} (1986) 99;
          Z. Phys. {\bf C 30} (1986) 279;
          Sov. J. Phys. {\bf 43} (1986) 450.
       \bibitem{sf}
           V. M. Belyaev, B. L. Ioffe and Ya. I. Kogan,
	Phys. Lett. {\bf 151 B} (1985) 290;\\
	S. Gupta, M. V. N. Murthy and J. Pasupathy,
	Phys. Rev. {\bf D 39} (1989) 2547;\\
	E. M. Henley, W-Y. P. Hwang and L. S. Kisslinger,
           Phys. Rev. {\bf D 46} (1992) 431;\\
           Ya. Ya. Balitskii, V. M. Braun and A. V. Kolesnichenko,
           JETP Lett. {\bf 50} (1989) 61;
           Phys. Lett. {\bf 242 B} (1990) 245; {\bf 318 B} (1993) 648 (E).
      \bibitem{m2-exp}
         P.J.Sutton, A.D. Martin, R.G. Roberts and W.J. Stirling,
         Phys. Rev. {\bf D 45} (1992) 2349.
      \bibitem{m2-lattice}
        G. Martinelli and C.T. Sachrajda,
        Phys. Lett. {\bf 196 B} (1987)184;
        Nucl. Phys. {\bf B 306} (1988) 865.
\end{thebibliography}
\end{document}